%% file: 0nuDBD-v4-pdflatex.tex
\documentclass[final,5p,times,twocolumn]{elsarticle}

\usepackage{amsmath}
\usepackage{amssymb}
\usepackage{graphicx}
\usepackage{multirow}
\usepackage{lineno}
\usepackage{varioref} % Provides \labelformat, which changes how~\ref references look
\usepackage[%
    pdftitle={CUORICINO analysis},%
    colorlinks=true,%
    linkcolor=blue,%
    citecolor=blue,%
    urlcolor=blue%
]{hyperref} % Load hyperref at the end of all the other packages but prior to other settings
\usepackage[all]{hypcap} 

\labelformat{equation}{\textup{(#1)}}
\labelformat{enumi}{\textup{(#1)}}

\def\ca{$\sim$}
 
\def\BBz{$0\nu\beta\beta$~}
\def\BBzn{$0\nu\beta\beta$}
\def\BBdn{$2\nu\beta\beta$}

\def\BB{$\beta\beta$~}
\def\BBn{$\beta\beta$}
\def\tect{$^{130}$Te~}
\def\tectn{$^{130}$Te}
\def\teod{TeO$_2$~}

\begin{document}
\title{$^{130}$Te Neutrinoless Double-Beta Decay with CUORICINO}

\input authors-CUORICINO.tex

\date{\today}

\begin{abstract}
We report the final result of the CUORICINO experiment. Operated between 2003 and 2008, with a total exposure of 19.75 kg$\cdot$y of \tectn, CUORICINO was able to set a lower bound on the \tect \BBz half-life of $2.8 \times 10^{24}$~years at 90\%~C.L. The limit here reported includes the effects of systematic uncertainties that are examined in detail in the paper. The corresponding upper bound on the neutrino Majorana mass is in the range 300--710 meV, depending on the adopted nuclear matrix element evaluation.
\end{abstract}

\maketitle

\section{Introduction}

\begin{figure*}[ht]
\label{fig:spectrum}
\begin{center}
\includegraphics[width=0.97 \linewidth]{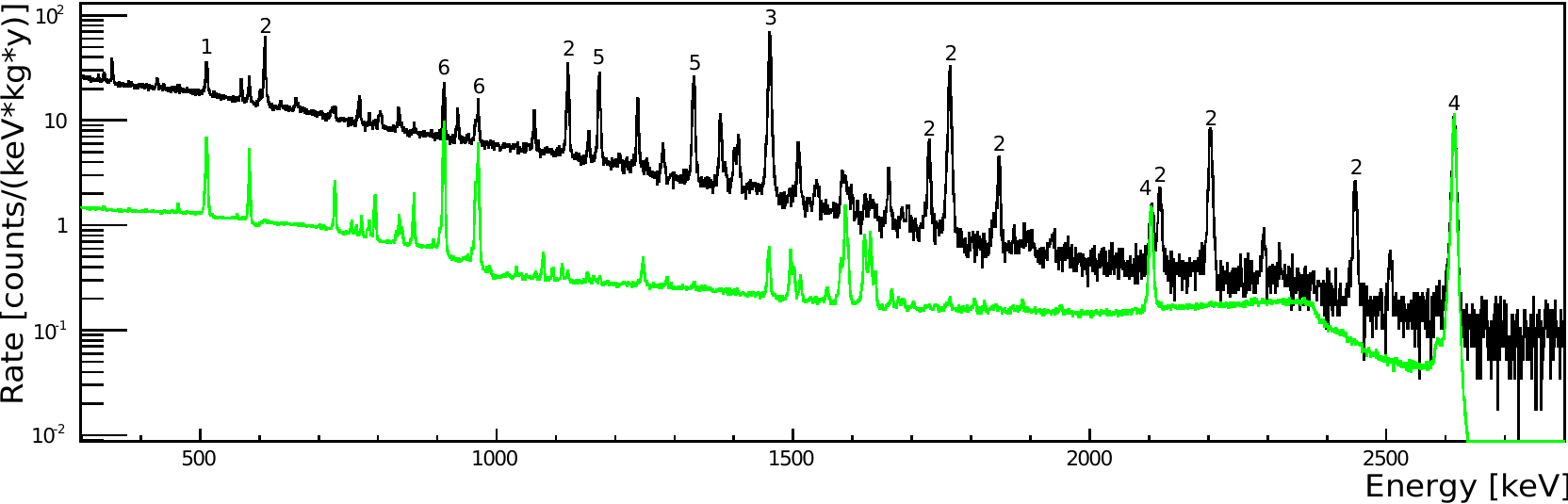} 
\end{center}
\caption{Anticoincidence total energy spectrum of all CUORICINO detectors (black). The most prominent peaks are labeled and come from known radioactive sources such as: $e^+e^-$ annihilation (1), $^{214}$Bi (2), $^{40}$K (3), $^{208}$Tl (4), $^{60}$Co (5) and  $^{228}$Ac (6). The total energy spectrum of all CUORICINO detectors during calibration measurements is also shown (color). For convenience, it is normalized to have the same intensity of the 2615~keV line of $^{208}$Tl as measured in the non-calibration spectrum.}
\end{figure*}

Double beta decay (\BBn) is a rare transition between two isobars, involving a change of the nuclear charge Z by two units. There exist several naturally-occurring even-even nuclei for which this is the only allowed decay mode. While the transition involving 2 electrons and 2 (anti)neutrinos (\BBdn) is allowed in any theoretical model (and has been observed for various isotopes), this is not the case for the neutrinoless channel (\BBzn). Despite being energetically possible, \BBz violates lepton number conservation by 2 units and is possible only if the neutrino is a massive Majorana particle~\cite{reviewDBD}.
\BBz searches have been pursued for more than half a century and today they experience a renewed interest, thanks to the discovery of neutrino oscillations ~\cite{Vissani, Fogli, Petcov}. Neutrino oscillations imply that neutrinos have a finite mass, but are not enough to determine the nature of that mass (Dirac or Majorana) or why it is so extraordinarily small.  Several theoretical speculations point toward a mass generation mechanism that implies a Majorana character of neutrinos, and that indicates the \BBz process as the unique tool with a discovery potential \cite{reviewDBD,Vissani}.

The \BBz transition could proceed through several different mechanisms, of which the simplest and most commonly cited is the exchange of a light Majorana neutrino. In this case the observation of \BBz would not only provide evidence for lepton number violation and the Majorana character of the neutrino, but would also result in a measurement of the effective Majorana mass $m_{ee}= \vert \sum m_i U_{ei}^2 \vert$ (where $m_i$ are the three mass eigenstates of neutrinos and $U_{ei}$ are the PMNS matrix elements) through the relation:

\begin{equation}
\frac{1}{\tau_{1/2}^{0\nu }} = m_{ee}^2 F_N = m_{ee}^2 G^{0\nu}\vert M^{0\nu} \vert ^2
\label{eq:vita} 
\end{equation}

\noindent Here $\tau_{1/2}^{0\nu }$ is the \BBz half-life, $G^{0\nu}$ is the two-body phase-space factor and  $M^{0\nu}$ is the \BBz nuclear matrix element (NME). The product $F_N = G^{0\nu}\vert M^{0\nu} \vert ^2$ is called the ``nuclear factor of merit"; the name refers to the fact that, according to equation~\ref{eq:vita}, $F_N$ directly influences the experimental sensitivity to $m_{ee}$. The main uncertainty in deriving $m_{ee}$ (or an upper limit on it) from the experimental result on $\tau_{1/2}^{0\nu}$ comes from the NME, which is a theoretical calculation still affected by a large spread among the adopted nuclear models and their implementations~\cite{NME1,NME2,NME3,NME4}. To mitigate this uncertainty, several candidate \BBz isotopes could be studied and the $m_{ee}$ results compared. Although it may not be feasible to study all the \BB candidates with reasonable sensitivity, there exists a ``golden list" of isotopes that -- as a compromise of cost, availability, technological approach and other factors -- have been studied so far. Among them, those that have yielded the most stringent limits on $m_{ee}$ (within the NME spread) are $^{76}$Ge~\cite{HM,IGEX}, $^{100}$Mo~\cite{NEMO}, $^{130}$Te~\cite{paperFrank} and $^{136}$Xe~\cite{Rita}. In all but one~\cite{Klapdor} case, only upper bounds on the Majorana mass have been reported.

In this paper, we discuss the final \BBz result of CUORICINO, which yields the most stringent bound on $m_{ee}$ based on \tect studies, and one of the best in general.  
CUORICINO data acquisition started in April 2003 and ended in June 2008. The data are separated into two runs (RUN I and RUN II), due to a major maintenance interruption. The data collection is summarized in table~\ref{tab:exp}. A partial data-set of~11.83 kg$\cdot$y of \tect exposure was used for the analysis reported in~\cite{paperFrank}.

The paper is organized as follows: after a short description of the experimental set-up in section \ref{sec:detector}, we present details of RUN II data analysis, discussing processing in section~\ref{sec:DA}, data reduction in section~\ref{sec:DR} and efficiency evaluation in section ~\ref{sec:efficiencies}. In section~\ref{sec:limit}, we describe two Bayesian approaches used for the \BBz half-life limit evaluation (one of them is that adopted in~\cite{paperFrank}), testing the two procedures on a toy Monte Carlo and discussing their compatibility on real data.  In section~\ref{sec:syst} we discuss the influence of systematic errors on the final result.

We conclude the paper with the CUORICINO final result for the \BBz half-life limit evaluated on the entire data set. This is done treating RUN I and RUN II as two independent experiments whose likelihoods are combined. This choice was motivated by the difference in detector configuration between the two runs (increased number of active detectors, improved performance) and a presumable difference in background composition (due to detector exposure to air).

\section{CUORICINO}\label{sec:detector}

\begin{table*}[t]
\caption[]{CUORICINO crystal information and statistics. Crystal mass is the average measured mass for CUORICINO detectors.}
%\begin{ruledtabular} 
\begin{center}
\begin{tabular}{ccccc}
%\toprule
\hline\noalign{\smallskip}\hline
~~~Crystal Type~~~&~~~Crystal Mass~~~&  ~~~$^{130}$Te Mass~~~   &  ~~~ Exposure Run II ~~~  & ~~~  Exposure Run I ~~~ \\
 &  [g] & [g] & [kg($^{130}$Te)$\cdot$y] & [kg($^{130}$Te)$\cdot$y]\\
\noalign{\smallskip}\hline\noalign{\smallskip}
big & 790 & 217 & 15.80 & 0.94 \\
small & 330 & 91 & 2.02  & 0.094\\
$^{130}$Te-enriched & 330 & 199 & 0.75 & 0.145\\
\noalign{\smallskip}\hline\hline
%\bottomrule
\end{tabular}
\end{center}
%\end{ruledtabular} 

\label{tab:exp}
\end{table*}
 
For many years, the most sensitive \BBz results for \tect have been obtained using bolometric detectors.  A bolometer is a type of calorimeter operated at ultra-low temperature, in which the energy of incident radiation is converted to heat, raising the temperature of the detector's body.
The energy released in the detector is then determined by measuring the temperature increase, in our case by using a semiconductor thermistor \cite{NTD} whose resistance varies exponentially with temperature. 
In this kind of detector, the candidate isotope is contained within the active mass of the detector itself. CUORICINO bolometers contain the isotope $^{130}$Te (isotopic abundance 33.8\%) which is a \BB candidate with a rather favorable factor of merit\footnote{A second \BB candidate with high natural abundance is the isotope $^{128}$Te (isotopic abundance 31.7\%). This isotope is not as interesting as \tect because of its lower transition energy, which reduces the nuclear factor of merit and shifts the signal to a higher background region (therefore also lowering the achievable experimental sensitivity).}.  

About 85\% of the time (see section~\ref{sec:DR}), the two electrons emitted by $^{130}$Te \BBz would be fully contained within one crystal. The signature of the decay would therefore consist of a monochromatic peak in the energy spectra of the bolometers at an energy equal to the Q-value of the decay: 2527.518$\pm$0.013~keV~\cite{0nuQvalue}.  The difficulty of the experiment lies in the control and reduction of all the background events that could mimic such a signal. These can be non-particle signals, due to electronic or thermal noise, or particle signals, due to radioactivity and cosmic rays. The former are rejected on the basis of a pulse shape discrimination technique (see section~\ref{sec:DR}). The latter are controlled  during the experiment's design and construction by proper material selection, handling, and shielding~\cite{paperFrank,ArtChambery,ArtRadio}, and  --- at the stage of data-analysis --- by coincidence cuts (see section~\ref{sec:DR}). 

CUORICINO was the latest step in a long series of experiments of increasing mass, performance and sensitivity~\cite{riv0,array4,array20}. The next experiment to use this technique will be CUORE~\cite{CUORE}, which is presently under construction.

CUORICINO~\cite{paperFrank} was a tower array of 62 \teod crystals used as bolometric detectors. The array was operated underground, in a dilution refrigerator located at the Laboratori Nazionali del Gran Sasso (INFN - Italy), which provides an average coverage of 1400 m of rock (3650 m.w.e.). CUORICINO crystals can be divided into four main groups according to their mass and isotopic abundance. These are: 
\begin{itemize}
\item the ``big crystals" --- 44 bolometers, $5\times5\times5~$cm$^3$ in size and 790~g in mass;
\item the ``small crystals" --- 14 bolometers, $3\times3\times6~$cm$^3$ in size and 330~g in mass;
\item the ``$^{130}$Te-enriched crystals" --- 2 bolometers, $3\times3\times6~$cm$^3$ in size and 330~g in mass, grown with \tect enriched material;
\item the ``$^{128}$Te-enriched crystals" --- 2 bolometers, with the same size and mass of the $^{130}$Te-enriched ones but grown with $^{128}$Te enriched material. 
\end{itemize}
The $^{130}$Te-enriched crystals have a \tect content corresponding to an isotopic abundance of 75\%~\cite{array20}, while the \tect content of 128-enriched crystals is so low that they will not be considered for the \BBz analysis presented in this paper. 

A detailed description of the array, cryogenic set-up, shields, front-end electronics, and DAQ can be found in~\cite{paperFrank} and references therein. Here we will describe the main steps of data acquisition and data handling which are relevant for the discussion.

The output voltage of each detector was monitored by a constant fraction trigger. When the output voltage exceeded the trigger threshold, the acquisition system recorded 512 samples (a 4 s window sampled at 125 Hz), which constitute one ``event". The acquired time window fully contained the pulse development, providing an accurate description of its waveform. A pre-trigger interval just prior to the onset of the pulse was used to measure the DC level of the detector, which corresponds to the instantaneous detector temperature.
The pulse amplitudes were evaluated offline for each recorded event, together with a few other characterizing parameters of the pulse. %The choice of this acquisition configuration was motivated by the amount of useful information that can be obtained by the recorded waveform.

Each single CUORICINO measurement lasted about 22 hours on average, with the time between measurements (about 2 hours) dedicated to cryogenic system maintenance. A routine calibration with an external $^{232}$Th source was performed approximately once per month, lasting for about 3 days. The accumulated data between two calibrations is referred to as a ``data-set". The spectrum obtained by summing all the CUORICINO collected data (i.e. summing over detectors and data-sets) is shown in figure~\ref{fig:spectrum}. The background recorded by the detectors is clearly dominated, in this region, by gamma emissions due to radioactive contaminations of the detector and of the surrounding apparatus. The most intense gamma lines are listed in reference~\cite{paperFrank}. In figure~\ref{fig:spectrum}, the spectrum corresponding to the sum of all calibration data is also shown. For convenience the calibration spectrum is normalized to have the same intensity of the 2615~keV line of $^{208}$Tl as measured in the background spectrum. \\

\begin{figure}[t]
\begin{center}
\includegraphics[width=1\linewidth]{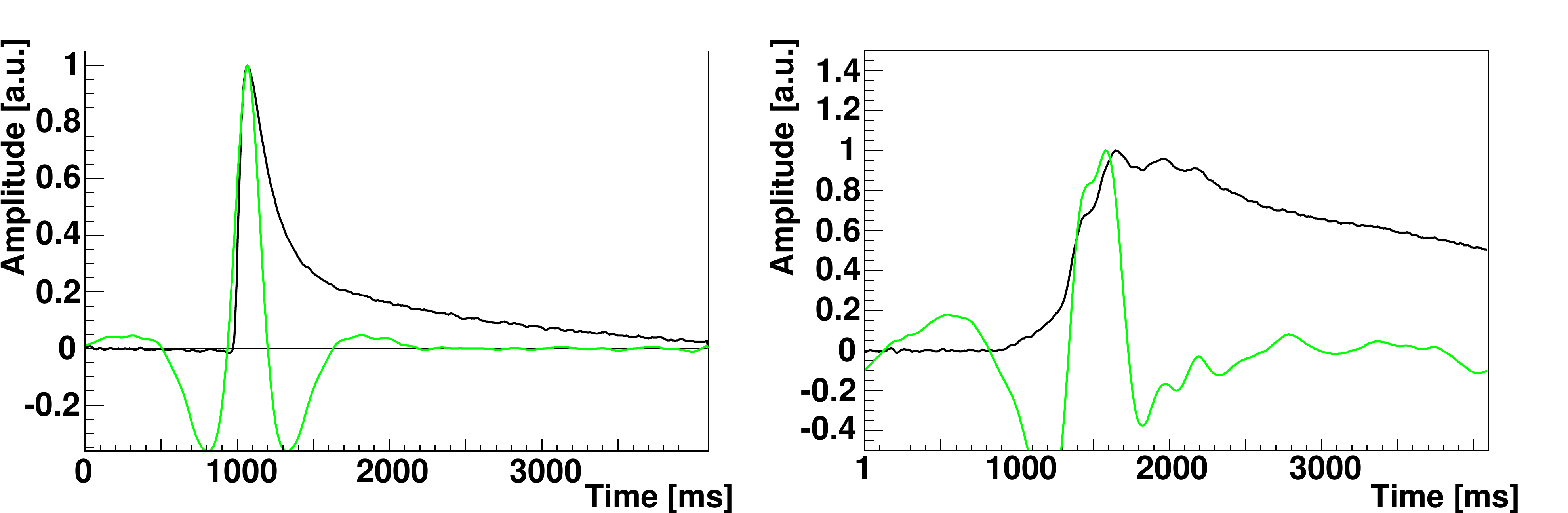} 
\end{center}
\caption{A bolometric particle event (left) and a spurious non-physical signal (right), superimposed with their optimum filter output (in color) in the time domain.}
\label{fig:spurious}
\end{figure}

\section{Data processing}\label{sec:DA}

The analysis of CUORICINO data starts with the collection of all the triggered events. For clarity, we will model the single waveform $V(t)$ induced by a particle interaction in the crystal as: 

\begin{eqnarray}
V(t) & = & V_0\,s(t)\,+\,n(t) \\
V_0 & =  &G(T) \cdot A(E)
\end{eqnarray}

In the first equation, $V_0$ is the maximum value of the raw signal acquired at time $t_0$, $s(t)$ describes the shape of the particle signal and $n(t)$ is an additive noise source. The second equation describes the dependence of the signal amplitude on the detector working temperature. Here we assume that the dependence of the gain on temperature, $G(T)$, and that of the amplitude on energy, $A(E)$, can be factorized. This is not true in general, however it is a good approximation when dealing --- as in our case --- with small temperature drifts. Although it describes a naive model, this formula highlights the key points of the analysis. In order to estimate $E$, we need:

\begin{enumerate} 
\item a technique able to measure $V_0$, reducing the effect of $n(t)$ as much as possible, in order to improve our resolution (amplitude evaluation);
\item an algorithm to control for the variation of $G(T)$ produced by detector temperature drifts (gain instability correction);
\item a technique to measure the form of $A(E)$ (energy calibration).
\end{enumerate}

\textbf{Amplitude evaluation}. This is done by maximizing the signal-to-noise ratio by means of optimum filtering~\cite{GattiManfredi}: each waveform is convolved with a transfer function $h(t)$ whose Fourier transform is defined as:  

\begin{equation}
H(\omega)=\mathrm{e}^{i\omega t_{max} } \frac{S^{*}(\omega)}{N(\omega)}
\end{equation} 

\noindent where $S(\omega)$ is the Fourier transform of the average detector response function $s(t)$, $N(\omega)$ is the spectral power density of the noise characterizing the detector, and $t_{max}$ is the time at which the pulse reaches its maximum. The functions $s(t)$ and $N(\omega)$ are computed by an averaging procedure of the bolometric pulses and of the Fourier transformed baselines\footnote{At random times during the course of data acquisition, sets of 512 samples (``baselines") were collected in anticoincidence with the trigger. These were used for the evaluation of the average noise power spectrum, as described in the text.}.

Figure~\ref{fig:spurious} shows an example of an event due to a particle interaction and and example of a non-particle event, most likely due to an abrupt temperature increase produced by an electric disturbance or by vibrations. Each waveform is superimposed with its optimum-filtered counterpart.\\
Once the optimum filter is applied, the amplitude of the signal is inferred from the maximum of the filtered waveform in time domain.\\

\textbf{Gain instability correction}. This correction is achieved by measuring the voltage amplitude, $V_{ref}$, of a monochromatic reference pulse.   
This pulse is produced by depositing a fixed amount of energy into the
crystal by the Joule dissipation from a heavily doped silicon resistor glued to the crystal.  Because the energy deposited is fixed, any variation of $V_{ref}$ would be due to a variation in $G(T)$, which can be measured and used to correct the amplitudes of all the triggered events. For a more detailed discussion of this method we refer to~\cite{pulser,stabilization}.\\

\begin{figure}[t ]
\begin{center}
\includegraphics[width=1\linewidth]{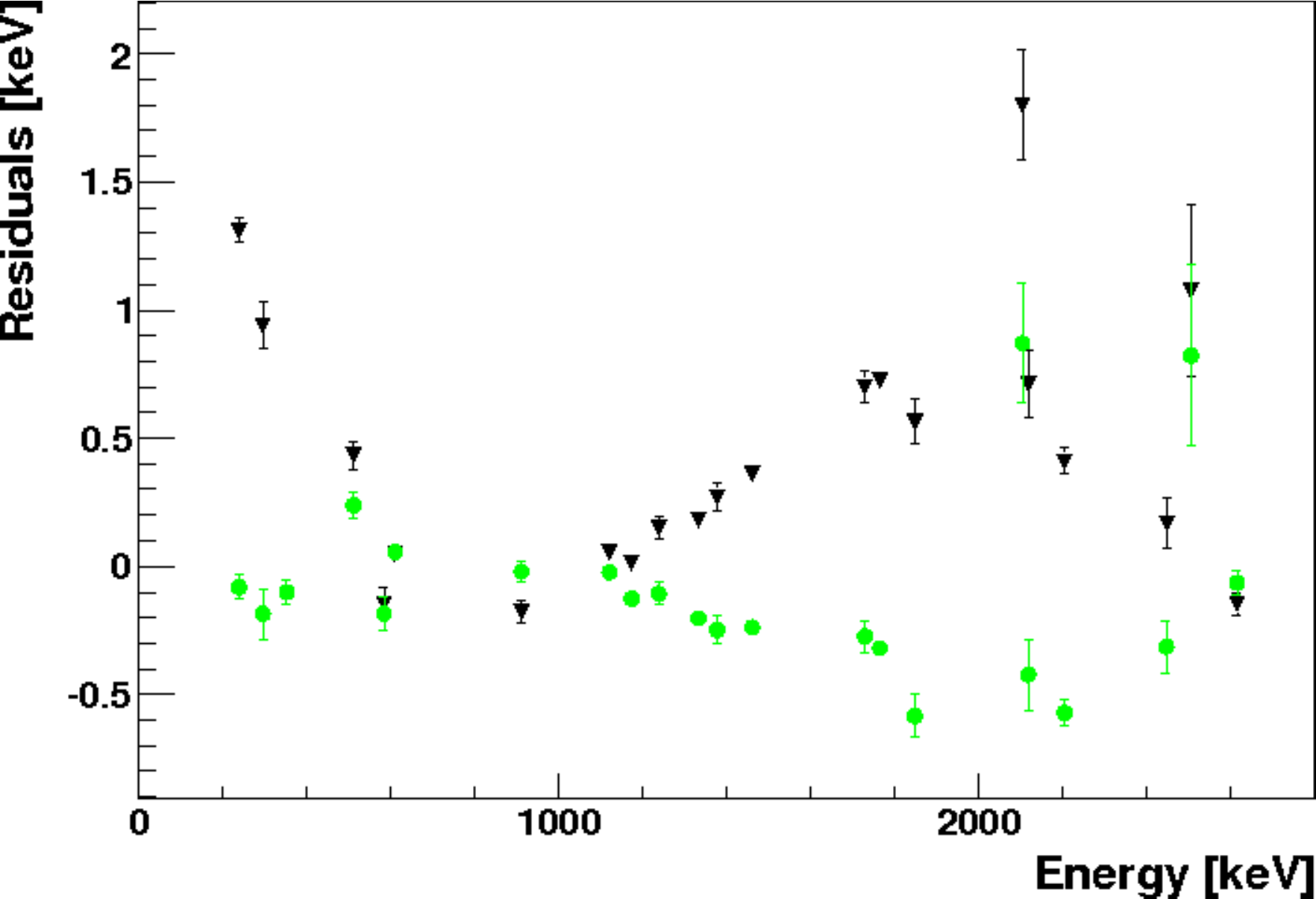} 
\caption{Residuals (nominal energy -- calibrated energy) vs. nominal energy evaluated on the main gamma lines identified in the CUORICINO background spectrum. Circles (in color) refer to a calibrated energy obtained with the third order polynomial, triangles (in black) with the log-polynomial. }
\label{fig:residuals}
\end{center}
\end{figure}

\begin{figure}[t floatfix]
\begin{center}
\includegraphics[width=0.8\linewidth]{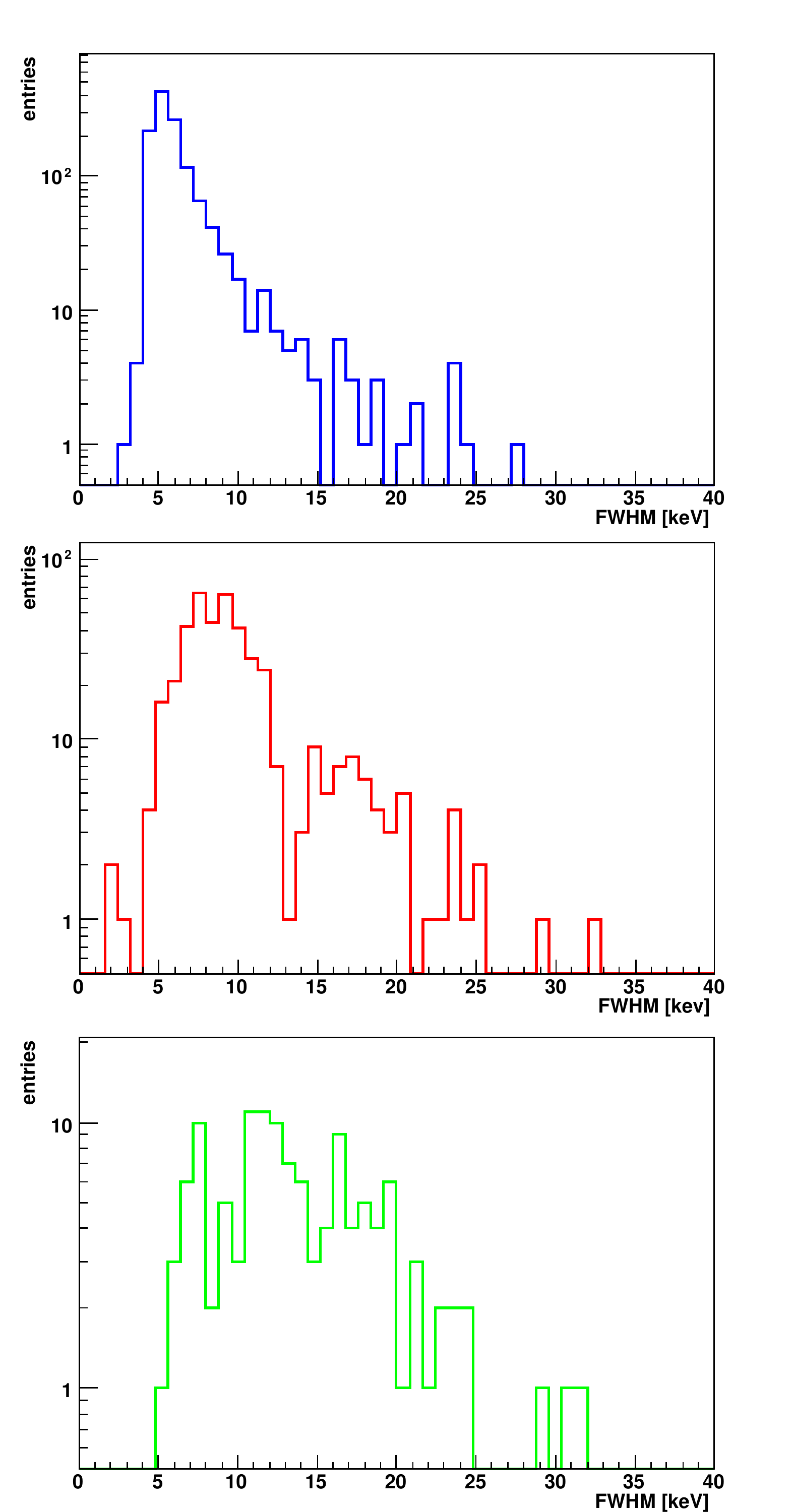}
\end{center}
\caption{Distribution of the energy resolutions (FWHM) measured in calibration for the three groups of crystals during the 33 data-sets belonging to RUN II. From top to bottom: big crystals, small natural crystals and \tect enriched crystals.} 
\label{fig:FWHMS}
\end{figure}

\textbf{Energy calibration}. The voltage-energy relationship is reconstructed by means of routine source calibrations: two wires of thoriated tungsten are periodically inserted between the cryostat and its external lead shield. The voltage amplitude of the pulses corresponding to the main gamma lines of $^{232}$Th are used for the determination of the parameters describing the $A(E)$ relationship. This function is characterized by different non-linearity sources~\cite{paperFrank}, the dominant one being the dependence of the thermistor resistance on the temperature~\cite{Vignati}. In this work, $A(E)$ is parametrized with a third-order polynomial, which can be considered as the truncated Taylor's expansion of the real unknown calibration function. In the previous CUORICINO analysis~\cite{paperFrank}, a different calibration function was used. This was a second order polynomial in $\textrm{log}(V)$ and $\textrm{log}(E)$, based on a thermal model describing our bolometric detectors. While this function performs better at extrapolation (i.e. above the highest calibration line at 2615~keV), the third order polynomial performs better in the interpolation region (i.e. between threshold and 2.6~MeV). The difference between the two parameterizations was studied using the total background spectrum recorded by CUORICINO (figure~\ref{fig:spectrum}); this spectrum contains several gamma peaks whose origins, and therefore nominal energies, are clearly identified. The difference between the nominal energy of each peak and its measured position (the residual) is plotted against the nominal energy in figure~\ref{fig:residuals}, showing the slightly better performance of the 3rd order polynomial. These residuals also provide important information concerning the precision of our calibration: their spread can be used as an estimator of the uncertainty in the energy position of a peak, including that produced by the \BBz signal.

Source calibration measurements are repeated for each data-set and are also used to check the detector performances over time. Figure~\ref{fig:FWHMS} shows the distribution of all the resolutions measured in calibration for the three crystal groups (big, small, and enriched crystals).

\section{Data reduction}
\label{sec:DR}
The final CUORICINO spectrum is composed of events which survived two different types of data selection: global and event-based cuts.\\

\textbf{Global cuts}: these are applied following quality criteria decided a priori (e.g. an excessive noise level or an incompatibility between the two calibration measurements at the beginning and the end of a data-set). They identify bad time intervals to be discarded. This kind of cut introduces a dead time that is accounted for by properly reducing the live time of the detector of interest. The cuts are generally based on off-line checks that monitor the detector performances and flag excessive deviations from global control quantities (average resolution, average rate, etc.).
The total dead time introduced by these global cuts is \ca5\%.  
A further dead time is introduced by the rejection of a short time window centered around each reference pulse (the frequency with which the reference pulses are generated is about 3 ~mHz). This cut ensures the rejection of possible pile-up of a particle signal with the reference pulse (the impact of this cut is reported as an efficiency in table~\ref{table:efficiencies}).\\

\textbf{Event-based cuts}: these are the pulse-shape and the anti-coincidence cuts. The former is used to reject non-physical and pile-up events (the presence of a pile-up prevents the optimum filter algorithm from providing a correct evaluation of the pulse amplitude). The latter allows for the reduction of the background counting rate in the region of interest (ROI). The \BBz signature we look for consists of a ``single-hit" event (only one detector at a time involved), while many of the background counts in the ROI are due to ``multiple-hit" events. These include events due to alpha decays on the crystal surfaces that deposit energy in two neighboring crystals and events due to gammas that Compton scatter in one crystal before interacting in another one.

\begin{figure}[tb]
\begin{center}
\includegraphics[width=1\linewidth]{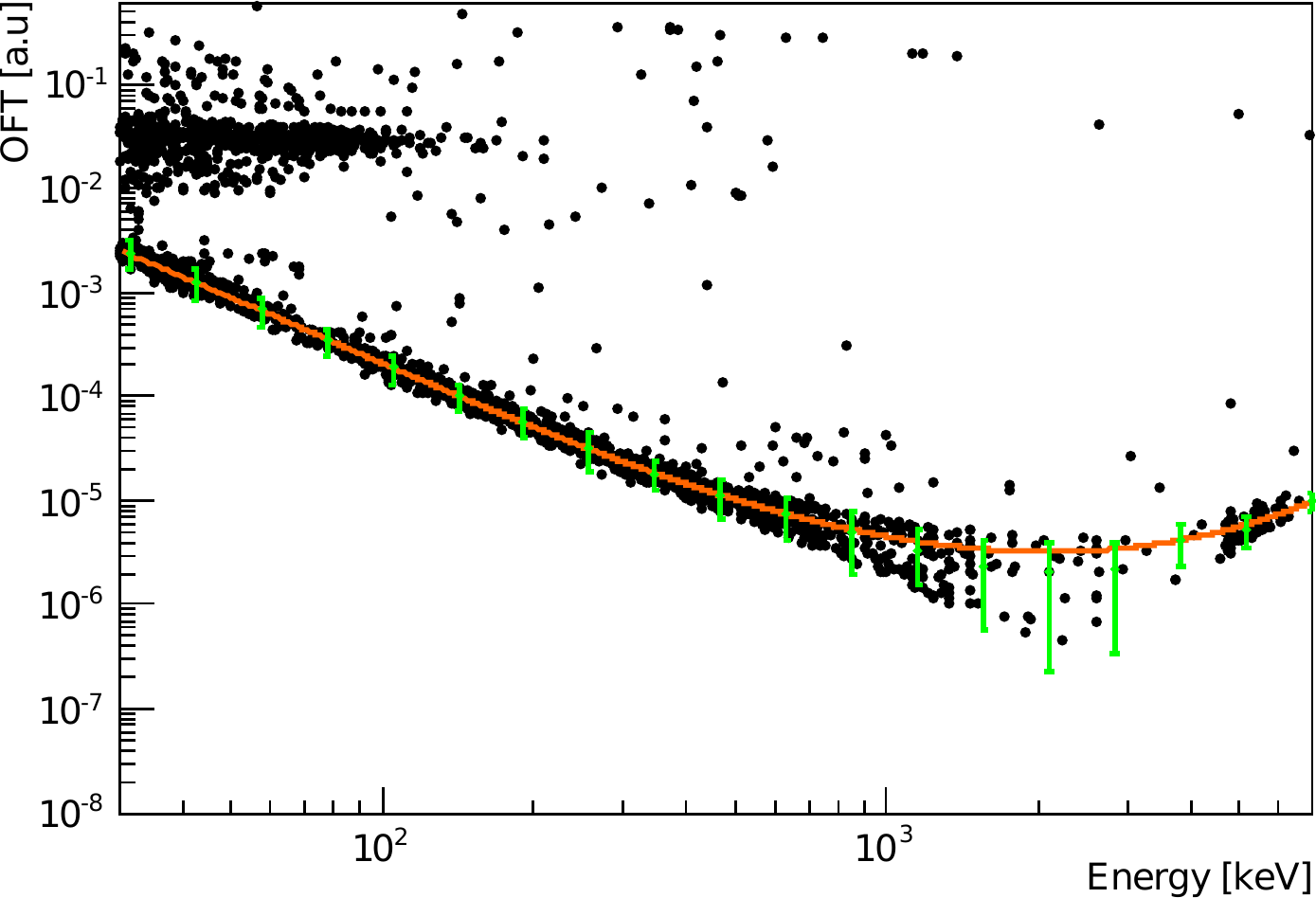} 
\end{center}
\caption{Typical scatter plot of the OFT (deviation of filtered raw signal from the average detector response) as a function of the signal's evaluated energy. The main trend identifies ``good" events. Pile-up and non-physical pulses are outside the confidence regions (identified by the colored vertical bars). At low energy, there is a high density distribution of events for OFT values between 10$^{-2}$ and 10$^{-1}$; these events are due to electric disturbances.}
\label{fig:psa}
\end{figure}

The pulse shape parameters used in this analysis are the rise time and decay time of the raw waveform and a parameter that measures the consistency of one of the basic statements of optimum filter theory.
This ``Optimum Filter Test" parameter (OFT) is the difference (expressed in percentage of the total amplitude) between the 
evaluation of the pulse height in the time domain (as the maximum value of the filtered pulse) and that in the frequency domain (as the integral of the filtered-pulse power spectrum). 
Indeed, if the shape of an event is identical to the average detector response, the two methods yield the same result. However, if the shape of the signal is different from the expected one (such as in the case of a non-physical or a pile-up event), they differ. Figure~\ref{fig:psa} shows the scatter plot of OFT as a function of energy for a CUORICINO detector (here only one data-set is reported). The main trend reflects the change of the signal shape with energy, and has a minimum in the region where the average response was measured (1-2 MeV). This variation in the pulse shape is caused by non-linearities introduced by the thermistor. Outliers on this plot correspond to misshapen events which will be discarded; the colored vertical bars identify confidence regions evaluated automatically on this distribution in order to obtain cuts which are independent of the signal amplitude.

The anticoincidence cuts require that only one detector fires within a time window of 100 ms. %Pulse shape and anti-coincidence cuts are connected since the definition of coincidences must consider only physical events.

\section{Signal efficiency} %ADAM
\label{sec:efficiencies}
\begin{table}[t]
\caption[]{Contributions to the CUORICINO \BBz signal efficiency.}
\begin{center}
\begin{tabular}{c r@{$\pm$}l}
\hline\noalign{\smallskip}\hline
~~~~ Source   ~~~~ &  
\multicolumn{2}{c}{~~~Signal efficiency (\%)~~~}\\ 
\noalign{\smallskip}\hline\noalign{\smallskip}

		 \multirow{2}*{Energy escape}  & 87.4 & 1.1 (big crystals)\\
		 & 84.2 & 1.4 (small crystals)\\
		 \noalign{\smallskip}\noalign{\smallskip}\noalign{\smallskip}
		 Pulse-shape cuts  & 98.5 & 0.3\\
		 \noalign{\smallskip}
		 Anti-coincidence cut & 99.3 & 0.1\\
		 \noalign{\smallskip}
		 Noise & 99.1 & 0.1\\
		 \noalign{\smallskip}
		 Pile-up with reference pulses  ~~& \multicolumn{2}{l}{97.7} \\
%		 \multirow{2}{Total} 
\noalign{\smallskip}\noalign{\smallskip}\noalign{\smallskip}
\multirow{2}*{Total}  & 82.8 & 1.1 (big crystals)\\
		 & 79.7 & 1.4 (small crystals)\\
		 \noalign{\smallskip}
\hline\noalign{\smallskip}\hline
\end{tabular}
\end{center}
\label{table:efficiencies} 
\end{table}

The signal efficiency is the probability that a \BBz event is detected, its energy is reconstructed accurately, and that it passes the data selection cuts. This parameter must be accurately determined, since it is used to obtain the number of \tect \BBz events. The overall signal efficiency is: (82.8 $\pm$ 1.1)\% for the CUORICINO big crystals and (79.7 $\pm$ 1.4)\% for the small and the $^{130}$Te-enriched ones. These efficiencies were computed as discussed below.

There are two main sources of inefficiencies, one ``physical" that can be computed by simulations, and the other ``instrumental" that must be measured from the data. The mechanism of ``physical" efficiency loss is the escape of a fraction of the \BBz energy from the source crystal. Mechanisms  for the ``instrumental" efficiency loss are: the pulse-shape cut, the anti-coincidence cut and an incorrect assignment of the energy of the signal (mainly due to noise and pile-up). Their contributions to the total signal efficiency are summarized in table~\ref{table:efficiencies}.\\ 
 
\textbf{Physical efficiency:} the \BBz signature is a sharp peak centered at the transition energy (Q-value) of the decay. The peak is produced by \BBz decays fully contained within the source crystal. The containment probability was evaluated using a Geant4-based Monte Carlo simulation that takes into account all the possible energy escape mechanisms (i.e. electrons, X-rays or bremsstrahlung photons escaping from the source crystals). Since the escape probability depends on the crystal geometry, the efficiency is slightly different for the big and the small crystals (see table~\ref{table:efficiencies}).

\textbf{Instrumental efficiency:} this is the product of the pulse-shape cut, anti-coincidence and excess noise efficiencies. 
To evaluate the efficiency of the pulse shape cut, the background photopeak at 2615 keV due to $^{208}$Tl was used as a proxy for the \BBz peak. The 2615~keV peak was chosen because of its proximity to the \BBz energy and its relatively high intensity. In principle, an ideal pulse-shape cut should leave the main peak untouched and should only reduce the flat background. The area of the peak can then be computed in terms of the the total number of signal events ($N_{sig}$), the signal efficiency  ($\epsilon_{PS}$),  the total number of background events ($N_{bkg}$) and the background efficiency ($\epsilon_{bkg}$). A simultaneous fit was done on both the spectra of accepted and rejected events. The area of the peak in the accepted events spectrum is given by $\epsilon_{PS}\times N_{sig}$, while for the rejected events it is $(1-\epsilon_{PS})\times N_{sig}$. Similarly, the background yield for the accepted events is $\epsilon_{bkg}\times N_{bkg}$, and the background yield for the rejected events is $(1-\epsilon_{bkg})\times N_{bkg}$. By including $\epsilon_{PS}$ directly in the parametrization of the fit, correlations among the fit parameters are automatically taken into account when the error on $\epsilon_{PS}$ is calculated. The result is $\epsilon_{PS} = (98.5 \pm 0.3)\%$, and $\epsilon_{bkg}= (64 \pm 2)\%$. The pulse shape cut is clearly very powerful, rejecting approximately 36\% of the events in the continuum background while retaining 98.5\% of the signal events in the peak. The events discarded in this region are mainly pile-up with real or spurious signals.

To estimate the efficiency of the anti-coincidence cut, we used the same procedure but considered the only available high intensity peak that is produced by a nuclear decay with no detectable coincidence radiation. This is the 1460~keV gamma line emitted in $^{40}$K electron capture (the only coincident radiation, a 3 keV X-ray, is far below our threshold). Since events in this photopeak are single hits, their reduction after an anti-coincidence cut can be ascribed only to random coincidences. 

The last source of inefficiency is the loss of \BBz events due to excess noise which can distort the pulse shape and introduce an error in the reconstructed energy. If such an error is greater than the resolution, the event can be considered as lost in the continuum background. In order to estimate this efficiency we compare the number of reference pulses generated during the measurements (the signals used for the gain instability correction, see section~\ref{sec:DA}) with the number actually measured in the correct energy range.

\section{\texorpdfstring{\BBz}{DBD} analysis}

\label{sec:limit}

\begin{figure}[t floatfix]
\begin{center}
\includegraphics[width=1\linewidth]{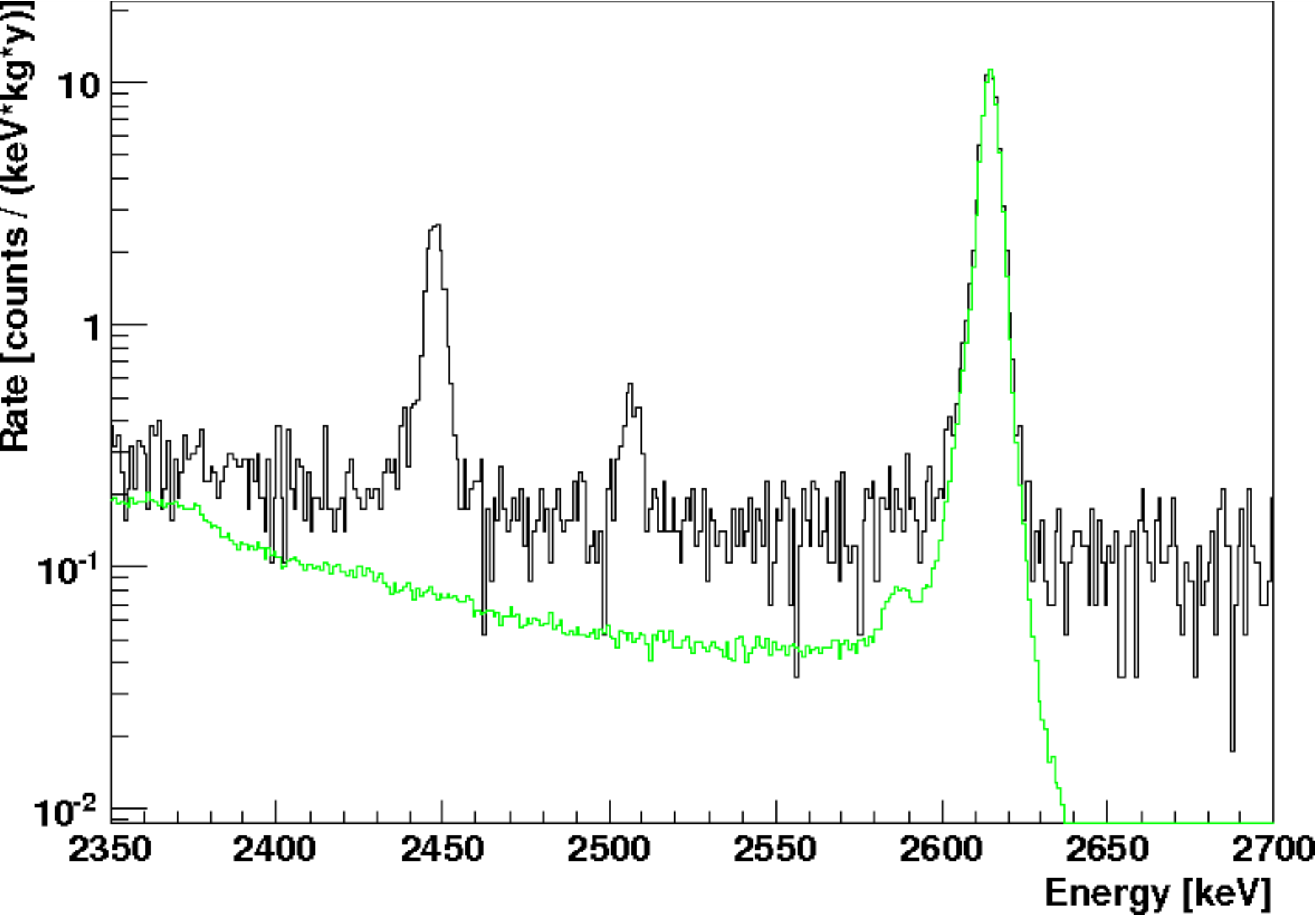} 
\caption{A closer view of the CUORICINO anticoincidence spectrum (presented in figure~\ref{fig:spectrum})  near the \BBz ROI. Note that the efficiencies listed in table~\ref{table:efficiencies} are not yet included.}
\label{fig:zoom}
\end{center}
\end{figure}

 The definition of the energy window used to fit the \BBz spectrum, the hypothesis assumed for the background shape and the number of free parameters used to describe the background itself are extremely important for the choice of the analysis procedure and for the determination of its systematics. The choice of the energy window is somewhat arbitrary, but it influences the background representation. If the energy window is too wide (compared to the signal FWHM) a very precise knowledge of the background shape is necessary. Obviously there is also a minimum width necessary to be able to evaluate the background level beyond the \BBz peak. In our case, there is a background line near the \BBz energy, at \ca2505 keV, due to $^{60}$Co (sum of the two photons emitted in cascade by $^{60}$Co decay), which should also be included in the window. Given these considerations, our final choice for the fit window is 2474--2580 keV. This is the widest window centered on the \BB Q-value that allows the following two background peaks to be excluded from the fit: the 2448~keV line of $^{214}$Bi and the 2587~keV Te X-ray escape peak of the $^{208}$Tl line. The latter peak is clearly visible in the CUORICINO calibration spectrum shown in figure~\ref{fig:zoom}, although --- due to the lower statistics --- it is not visible in the background spectrum shown in the same figure.
 
 Within this picture, we choose the simplest  possible model for the shape $f_{i,j}(E)$ of the spectrum (normalized to mass, live time, efficiency and isotopic abundance) of each single detector (index $i$) in each data-set (index $j$) as: 

\begin{eqnarray} 
f_{i,j}(E) = B_i \,+\, \Gamma^{^{60}Co}_{i,j} \,\,g_{i,j}\left( E-E^{Co} \right) + \Gamma ^{0\nu} \,\,g_{i,j} \left( E-E^{0\nu}\right)
\label{eq:pdf}
\end{eqnarray}

\noindent Here $g_{i,j}(E)$ is the function describing the shape of monochromatic energy lines in the $i^{th}$ detector, during the $j^{th}$ data-set, i.e. the response function that is represented by a gaussian with a width determined from calibration data\footnote{To evaluate the energy resolution in the \BBz region we use the 2615~keV peak since this is the nearest peak to the Q-value clearly visible in our calibration spectra.}. $E^{0 \nu }$ is the \tect \BB Q-value, fixed at its measured value (2527.5~keV). $E^{Co}$ is the sum energy of the two $^{60}$Co gamma lines (2505.7~keV). $B_i$ is the flat background component for the $i^{th}$ detector (here we assume a time independent background). Finally $\Gamma^{^{60}Co}_{i,j}$  and $\Gamma^{0 \nu \beta \beta }$  are, respectively, the $^{60}$Co activity for the $i^{th}$ detector during the $j^{th}$ data-set ($^{60}$Co has a half-life of 5.27 years), and the absolute activity for \BBz, both expressed in counts/kg/y.

Free parameters are $B_i$, $\Gamma^{^{60}Co}_{i,j}$, $\Gamma ^{0\nu}$ and $E^{Co}$, the parameters of the response function being fixed at the values measured during calibrations. Note that the dependence of $\Gamma^{^{60}Co}_{i,j}$ on the index $j$ is determined by the $^{60}$Co half-life; therefore, the total number of free parameters is determined only by the number of detectors (i.e. by the index $i$).

\begin{figure}[t floatfix]
\begin{center}
\includegraphics[width=0.95\linewidth]{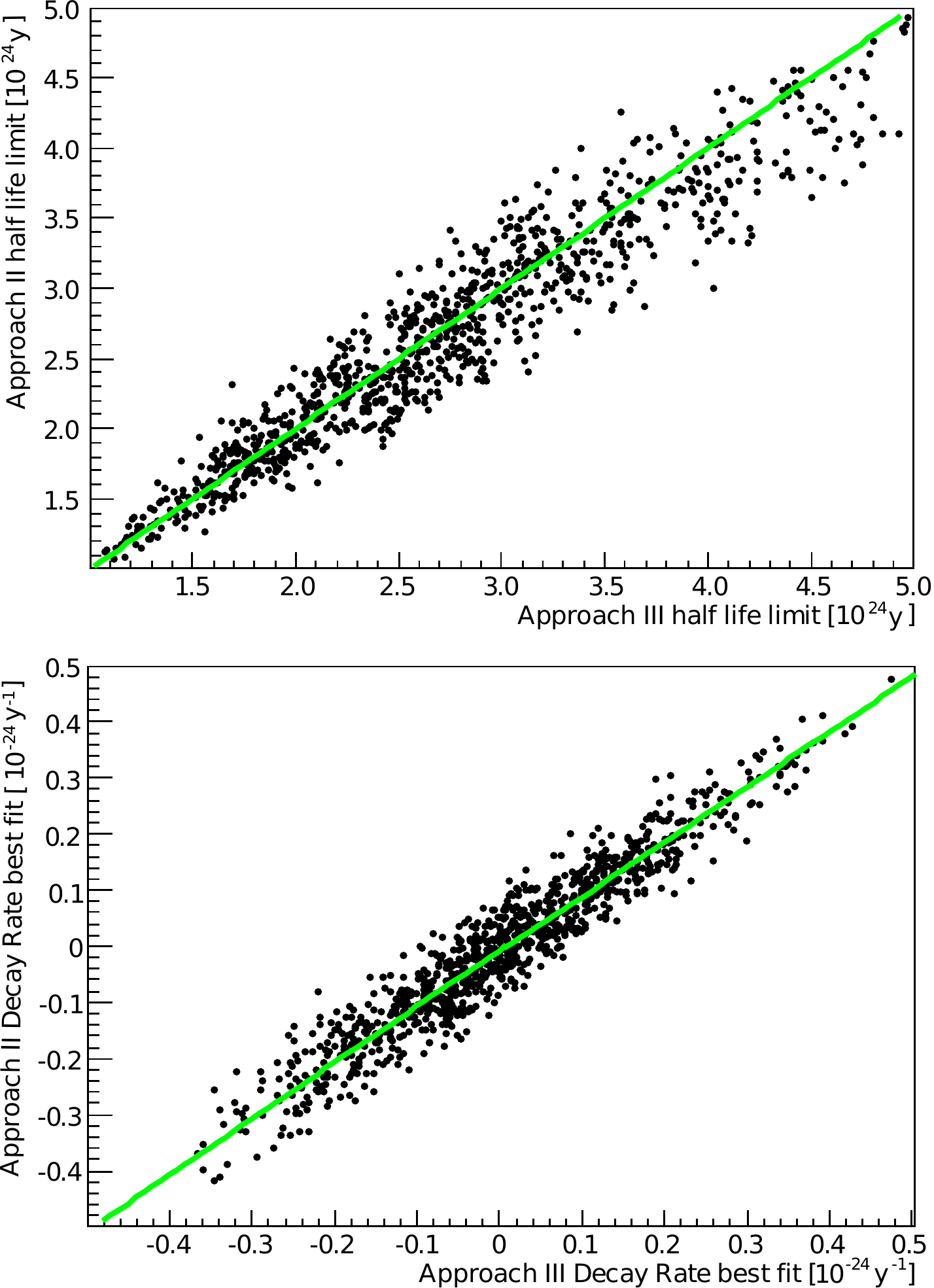}
\end{center}
\caption{Results of a toy Monte Carlo simulation with no \BBz signal (1000 simulated CUORICINO-RUN II experiments). Scatter plot of the 90\% C.L. limits (top panel) and of the best fits (bottom panel) obtained with the two different approaches. The colored line has slope=1 and shows the strong correlation between the two techniques.} 
\label{fig:limit_scatter}
\end{figure}

\begin{figure*}[t floatfix]
\begin{center}
\includegraphics[width=0.95\linewidth]{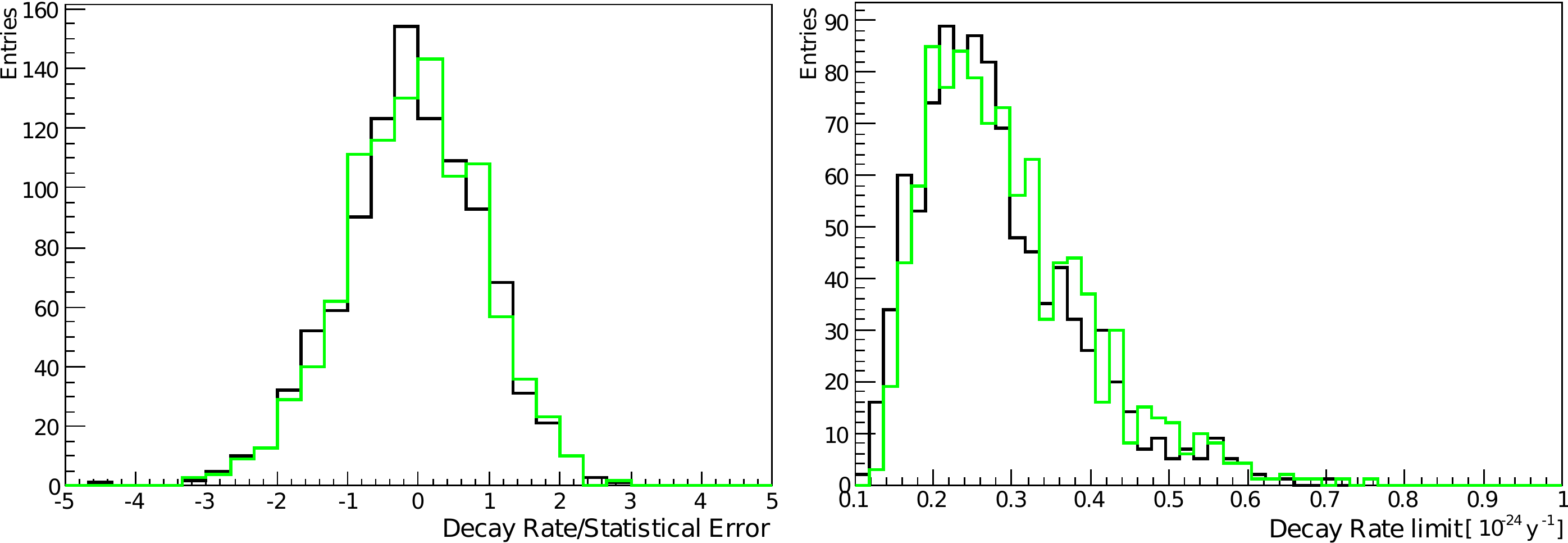}
\end{center}
\caption{Results of a toy Monte Carlo simulation with no \BBz signal (approach II in black and approach III in color). Left panel: pull distributions of the obtained best fits divided by their statistical error. Right panel: distribution of the 90\% confidence level limit on the decay rate.}
\label{fig:pull}
\end{figure*}

\subsection{Statistical approaches} 
\label{sec:limit_evaluation}
The CUORICINO spectrum shows no evidence of a \BBz signal, thus we will provide a limit for the half life of \tect by means of a Bayesian approach. Unlike other low statistics methods such as that of Feldman and Cousins~\cite{FC}, this technique does not require an exact evaluation of the expected number of background events (which is unknown). In our case, all the uncertainties are marginalized in the process of the limit computation, and our prior knowledge for the rate $\Gamma^{0\nu}$ will be represented by a flat distribution, excluding the non-physical region.

Once the statistical method is chosen, we need to decide how to model the experiment: every CUORICINO detector can in fact be imagined as an independent search for \BBzn, characterized by its own background and resolution. There are three natural approaches which can be chosen to search for a \BBz signal:\\
\begin{description}
\item I  \hskip1em--- treat the different detectors separately; 
\item II \hskip0.5em--- treat the different detectors separately, assuming an identical background for all detectors within each group (big, small, $^{130}$Te-enriched);
\item III \hskip0.2em--- sum the spectra of all detectors belonging to the same group.
\end{description}
In approach I, each detector and each data-set is fit with its own function $f_{i,j}(E)$. In principle this is the best approach since it uses all the information available; however, the number of free parameters is huge (about a hundred).

Approach II lowers the number of free parameters by forcing the background and the $^{60}$Co rates to be identical on detectors of the same group. In this approach each detector and data-set is still described individually by its own function $f_{i,j}(E)$ but the total number of free parameters is reduced since B$_i$ can assume only 3 values (for big, small and $^{130}$Te-enriched crystals) and the same is true for $\Gamma^{^{60}Co}_{i,j}$. This could be considered a strong assumption, but it is motivated by the fact that the low statistics prevent us from being sensitive to background variations among crystals of the same group in the \BBz region. This method also offers the advantage of being less sensitive to fluctuations in the counting rate of a single detector over time, and takes into account the decay rate of $^{60}$Co. 

Approach III removes the background assumption of the previous model, at the price of a certain degree of information loss. The counting rate is simply averaged over all data-sets and detectors for the three mentioned groups. A variation of the background and of the $^{60}$Co rate over time is then irrelevant, provided that the response function does not change with time. The average is done simply by summing over all the data collected with detectors belonging to the same group, thereby obtaining three spectra that can be represented by the function:

\begin{eqnarray} 
f_{k}(E) = B_k \,+\, \Gamma^{^{60}Co}_{k} \,\,G_{k}\left( E-E^{Co} \right) + \Gamma ^{0\nu} \,\,G_{k} \left( E-E^{0\nu}\right)
\label{eq:pdf2}
\end{eqnarray}

\noindent Here, the index $k$ has three allowed values for big, small and $^{130}$Te-enriched crystals while the response function  $G_k(E)$ is defined as: 
\begin{equation} 
G_k(E)=\frac{1}{\sum \limits_{i,j} A_{i,j}}\sum_{i,j} \frac{A_{i,j}}{\sqrt{2 \pi} \sigma_{i,j}}  \exp \left( -\frac{(E-E_0)^2}{2\sigma_{i,j} ^2} \right ) 
\label{eq:maria_function}
\end{equation}
\noindent  where the sum over $i$ extends on all the detectors belonging to the $k^{th}$ group and $j$ runs over all the data-sets. $A_{i,j}$  and $\sigma_{i,j}$ are the corresponding background exposure and energy resolution measured during calibration. Note that --- as is true also for the other two approaches --- the response function is built using measured quantities, i.e. it does not contain any free parameters.

% so that index $i$ in equation~\ref{eq:pdf} runs only on the crystal type (big, small and $^{130}$Te-enriched).  Equation~~\ref{eq:pdf} can still be used to represent the counting rate in the \BBz region of the three group of detectors if the response function is modified in order to take into account that within each group the different detectors have a different energy resolution. \\

We discarded the first approach due to the excessive number of free parameters, and we performed two parallel (and independent) analyses on real and Monte Carlo-simulated data for the limit computation, following an unbinned likelihood technique \cite{unbinned} for the second approach and the standard CUORICINO Likelihood-Chi-Square technique \cite{paperFrank,CUORICINO1} for the third one. The goal was to choose the most reliable procedure, checking for possible biases and comparing performances.

\begin{figure*}[t floatfix]
\begin{center}
\includegraphics[width=0.48\linewidth]{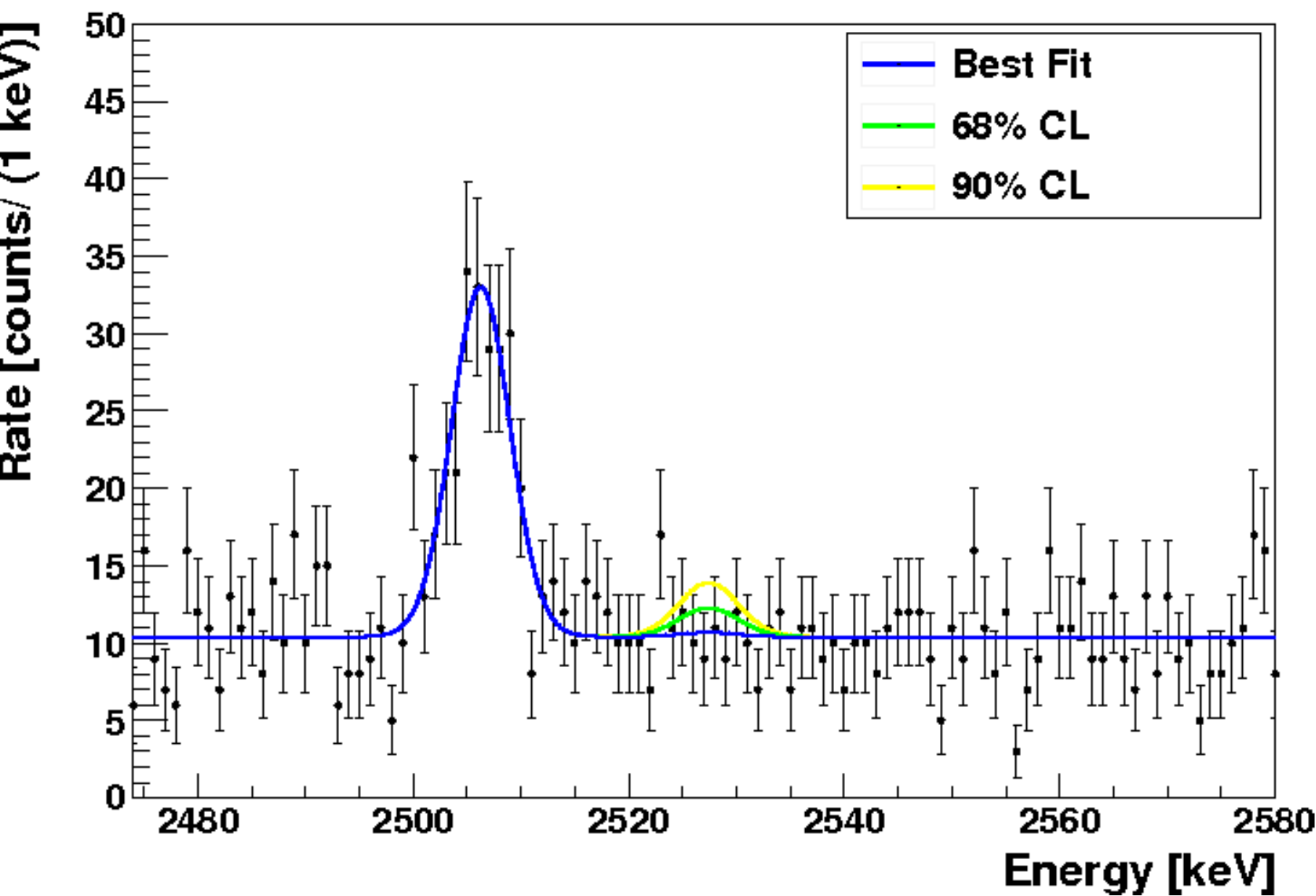}
\includegraphics[width=0.48\linewidth]{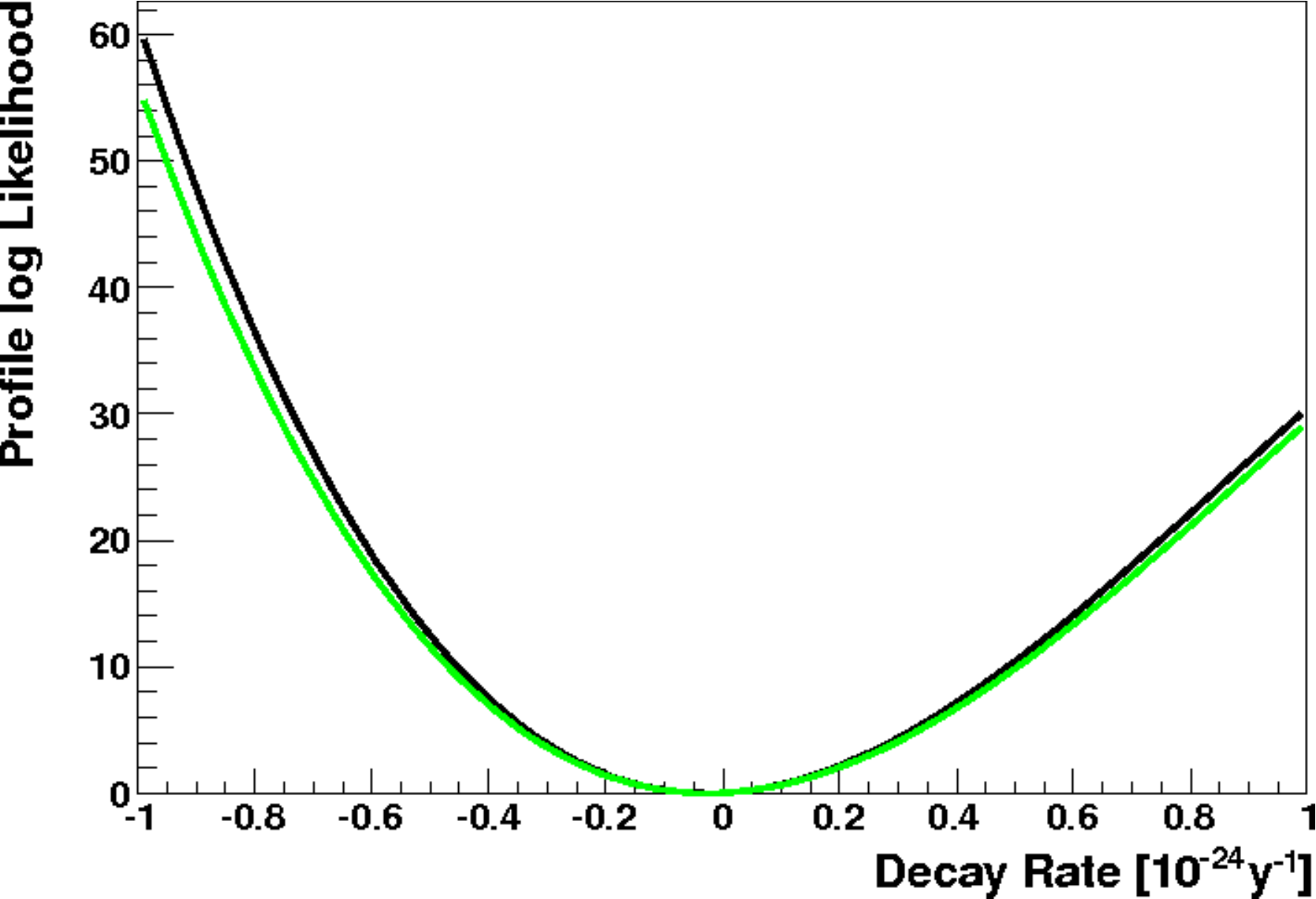}
\end{center}
\caption{Left panel: best fit, 68\% and 90\% confidence intervals for the total statistics (RUN I+RUN II) superimposed on the CUORICINO sum spectrum of the three groups of crystals (each scaled by efficiency and exposure) in the \BBz region. (The purpose of the plot is to give a pictorial view of the result; in fact the fit was performed separately on 6 spectra whose likelihood are combined, as described in the text). Right panel: negative profile of the combined log likelihoods of RUN I and RUN II before (black) and after (color) the systematic uncertainty is included.} 
\label{fig:DBDrate}
\end{figure*}

\subsection{Comparison of methods on real data and Monte Carlo simulations} 
\label{sec:toy}
%expected above the background level, it is mandatory to have a strong knowledge of the sensitivity of the experiment, even more than the limit itself which can be strongly affected by random noise fluctuations. Starting from the definition of the sensitivity given in F\&C as the average value of the limit among similar experiments with the same average background we constructed a method based on a toy Monte Carlo model of our experiment, in order to have a better picture of the impact of the fit method on the limit evaluation: CUORICINO 

\begin{table}[t]
\caption[]{Comparison between the fit results for the two studied approaches on RUN II data.} 
\begin{center}
\begin{tabular}{ccc}
\noalign{\smallskip}\hline\noalign{\smallskip}\hline
 & ~~Method II~~ & ~~Method III ~~\\
 \noalign{\smallskip}\hline\noalign{\smallskip}
Best Fit [y$^{-1}$]&  (0.2 $\pm$ 1.5)$\times$10$^{-25}$ & (0.3 $\pm$ 1.5)$\times$10$^{-25}$ \\
Half-life limit [y] & 2.5 $\times$ 10$^{24}$  & 2.4 $\times$ 10$^{24}$  \\
\noalign{\smallskip}\hline\noalign{\smallskip}\hline
\end{tabular}
\end{center}
\label{table:methods_results} 
\end{table}

Applying the two methods on the CUORICINO RUN II data, we obtained results which were indeed very similar. Table~\ref{table:methods_results} shows the best fits and limits for the two approaches: no significant difference is observed.

In order evaluate the performances of the two approaches, we compared the distribution of the relevant statistical estimators for a thousand toy Monte Carlo simulated spectra for each CUORICINO detector, generated with rates corresponding to those directly measured in RUN II (see table~\ref{table:lim_results_runI}) and no \BBz signal. Several pieces of information have been obtained from this comparison:
\begin{itemize}
\item Both methods lead to compatible results. Figure~\ref{fig:limit_scatter} demonstrates a strong correlation between them.
\item Both are unbiased. Figure~\ref{fig:pull} (left panel) shows that for both methods, the distribution of the best fits divided by their statistical errors is compatible with a gaussian centered at zero with a variance equal to one. This is an important result which is not always guaranteed by maximum likelihood methods applied to low statistics. 
\item Both have similar sensitivities. The distributions of the 90\% confidence intervals (figure~\ref{fig:limit_scatter} top panel) show a sensitivity of nearly 2.5x10$^{24}$~y, evaluated as the median of the distribution, following the Feldmann and Cousins prescription~\cite{FC}.
\end{itemize} 
The last point offers a nice synthesis for a better understanding of the numbers we are dealing with. A wide range of limits can be reached in experiments with the same true background level (figure~\ref{fig:pull}, right panel) and therefore with the same sensitivity. In this respect, quoting only the limit reached by an experiment can be misleading if the sensitivity is not also mentioned.

Having such an impressive correspondence between the two approaches, the choice of the one or the other is somewhat arbitrary. We opted for the third method because it is consistent with the previous analysis of CUORICINO data \cite{paperFrank,CUORICINO1} and because of its intrinsic simplicity.

\begin{table*}[]
\caption{Background and $^{60}$Co rates expressed in counts/keV/kg/y obtained by the combined fit of RUN I and RUN II data. The absolute rates are about 20\% higher because the efficiencies have not been yet included. The high background rate obtained for $^{130}$Te-enriched detectors is mainly due to a higher intrinsic contamination of these crystals.}
\begin{center}

\begin{tabular}{lccc}
\noalign{\smallskip}\hline\noalign{\smallskip}\hline
~~~Fit Parameter ~~~ &  ~~~big crystals~~~ & ~~~small crystals~~~ & ~~~$^{130}$Te-enriched crystals~~~\\
\noalign{\smallskip}\hline\noalign{\smallskip}
RUN \,I: flat background rate& 0.20$\pm$0.02 &0.20$\pm$0.02 &  0.8$\pm$0.4    \\
RUN II: flat background rate & 0.153$\pm$0.006 & 0.17$\pm$0.02 & 0.35$\pm$0.05 \\
RUN \,I: $^{60}$Co rate & 4.6 $\pm$ 1.5  & 9 $\pm$ 6 & 0 $\pm$ 14\\
RUN II: $^{60}$Co rate & 2.5 $\pm$ 0.3 & 1.7 $\pm$ 0.8 & 0 $\pm$ 3.5\\
\noalign{\smallskip}\hline\noalign{\smallskip}\hline
\end{tabular}
\end{center}
\label{table:lim_results_runI}

\end{table*}

\section{Systematic uncertainties} %ADAM
\label{sec:syst}
In our analysis we have identified the following sources of systematic uncertainties:
\begin{itemize} 
\item the calibration uncertainty;
\item  uncertainty in the signal efficiency;
\item the background shape;
\item  the energy window of the fit.
\end{itemize}
In the case of the calibration uncertainty we have a direct estimate of its magnitude coming from a dedicated analysis of the residuals (see section~\ref{sec:DA}): we reconstruct the position of a peak in the \BBz region with a precision $\Delta E=\pm0.8$~keV. This systematic error has been included directly in the fit as a gaussian fluctuation in the energy position. Since this uncertainty is significantly larger than the error (0.013 keV) on the Q-value~\cite{0nuQvalue}, the systematic error discussed here  also automatically includes the experimental uncertainty on the measured \tect transition energy.

To evaluate the uncertainty from the choice of energy window and shape of the background spectrum,  we varied the model for the background (flat, linear and parabolic) at four different energy intervals centered at the Q-value of the \BBz decay (we increased the lower and the upper bound of an energy window starting from 2527$\pm$30 keV at increasing steps of 5 keV). The results of this analysis have shown an average variation in $ \Gamma^{0\nu}$ of $3\times10^{-26}$y$^{-1}$.

The uncertainty on the signal efficiency is reported in table~\ref{table:efficiencies} to be 1.1\% for the big crystals and 1.4\% for the small crystals, both of which are negligible compared to the contributions from the energy scale and background parametrization uncertainties.

\label{sec:calunc}

\section{The CUORICINO final result}

\begin{table*}[ht]
\caption[]{We compare the most stringent 90\%~C.L. half-life lower limits present in literature (column 2). For each experimental result (rows 1-4) and for each of the considered NME evaluations (column 3-6), we report a $m_{ee}$ range. This identifies the upper bound on the neutrino Majorana mass according to the different results reported by the same author (when varying some of the parameters in the used nuclear model). In the two lower rows we compare the $m_{ee}$ range obtained for the 95\% C.L. half-life limit on \tect (row 5) with the positive signal quoted by ~\cite{Klapdor} (row 6). For this last case the $m_{ee}$ range corresponds to the 2 sigma range in the measured half-life. }
\begin{center}
\begin{tabular}{cccccc}
\noalign{\smallskip}\hline\noalign{\smallskip}\hline
~~~Isotope ~~~ &  ~~~$\tau_{1/2}^{0\nu }$~~~ & ~~~QRPA~\cite{NME1}~~~ & ~~~QRPA~\cite{NME2}~~~ & ~~~SM~\cite{NME3}~~~ & ~~~IBM~\cite{NME4}~~~\\
& [y] & $m_{ee}$ [meV] & $m_{ee}$ [meV] & $m_{ee}$ [meV] & $m_{ee}$ [meV] \\ 
\noalign{\smallskip}\hline\noalign{\smallskip}
\tect (CUORICINO, this work) & $> 2.8~10^{24}$ & $<$ 300--570 & $<$ 360--580 & $<$ 570--710 & $<$ 350--370\\
$^{76}$Ge (Heidelberg-Moscow collaboration~\cite{HM}) & $> 1.9 10^{25}$ & $<$ 230--400 & $<$ 280--460 & $<$ 530--640 & $<$ 270\\
$^{100}$Mo (NEMO collaboration~\cite{NEMO}) & $> 5.8~10^{23}$ & $<$ 610--1270 & $<$ 810--1430 & - & $<$ 830--850 \\
$^{136}$Xe (Dama/LXe~\cite{Rita}) & $> 1.2~10^{24}$ & $<$ 700--1640 & $<$ 800--1230 & $<$ 1020--1270 & $<$ 640--670\\
\hline
\tect (CUORICINO, this work, 95\% C.L.) & $> 2.3~10^{24}$ & $<$ 340--630 & $<$ 390--640 & $<$ 620--780 & $<$ 390--410\\
$^{76}$Ge (Heidelberg-Moscow experiment~\cite{Klapdor}) & $ =2.23^{+0.88}_{-0.62} 10^{25}$ & = 180--430 & = 220--500 & = 410--700 & = 210--290\\
\noalign{\smallskip}\hline\noalign{\smallskip}\hline
\end{tabular}
\end{center}
\label{table:NME}
\end{table*}

As a next step, we added the contributions from big, small and enriched crystals from RUN I, combining their likelihoods with the RUN II data and using a similar reconstruction for the response function as described in section~\ref{sec:limit_evaluation}.\\
The background rates are shown in table~\ref{table:lim_results_runI} while in figure~\ref{fig:DBDrate} (left panel) we show the best fit and the corresponding 68\% and 90\%~C.L. limits. Figure~\ref{fig:DBDrate} (right panel) shows the logarithm of the combined likelihoods of RUN I and RUN II  before and after the systematic uncertainties are included.\\

The resulting best fit for the \BBz rate of $^{130}$Te is: 
\begin{equation} 
\Gamma^{0\nu}_{\mathrm{best}}= (-0.25 \pm 1.44(\mathrm{stat}) \pm 0.3(\mathrm{syst})) \times10^{-25}\mathrm{y}^{-1} \nonumber
\end{equation} 
This result is compatible with zero, and the corresponding 90\%~C.L. half life lower bound is:
\begin{equation}
\tau_{1/2}^{0\nu }\geq 2.9\times10^{24} \mathrm{y}\nonumber
\end{equation} 

This limit is almost identical to the one we published in \cite{paperFrank}, despite the increase of the total exposure by a factor \ca1.6. In general it is expected that the limit scales with the square root of the exposure, i.e. we would expect  an improvement of about a factor 1.3 which is by far smaller than the spread in the 90\%~C.L. limits that different experiments (with the same exposure and sensitivity) can yield (figure~\ref{fig:pull}). This is the reason we prefer to quote the sensitivity of the experiment together with the limit. 
It should to be mentioned that, in reference \cite{paperFrank}, we used an older value of the \tect transition energy energy which had a much larger error and a slightly higher central value than the recently measured one \cite{0nuQvalue,Qv2}. With the same data used in \cite{paperFrank}, using the new result for the transition energy (with its smaller error) pushes the limit toward a lower half-life. 

The inclusion of the systematic error modifies the likelihood profile for our data as shown in the right panel of figure~\ref{fig:DBDrate}. The profile can be considered the $\chi^2$ of our fit as a function of all the possible $\Gamma^{0\nu}$. Thus we will refer to it as $\chi^2_{stat}$. If we adopt the hypothesis that our knowledge of $\Gamma^{0\nu}$ is smeared --- near the best fit values --- by a gaussian systematic uncertainty of magnitude $\sigma_{syst}$, the total $\chi^2_{tot}$ will be:
\begin{equation} 
\frac{1}{\chi^2_{tot}} = \frac{1}{\chi^2_{stat}}+\frac{1}{\chi^2_{syst}}
\end{equation} 

\noindent where the simplest approximated form of $\chi^2_{syst}$ is:
\begin{equation}
\chi^2_{syst} = \frac{\left(\Gamma^{0\nu} - \Gamma^{0\nu}_{\mathrm{best}}\right)^2}{\sigma_{syst}^2}
\end{equation} 

With this modification of $\chi^2$ and because the systematic uncertainty is small compared to the statistical error, we obtain a slightly weaker limit on the half life:

\begin{equation}
\tau_{1/2}^{0\nu }\geq 2.8\times10^{24} \mathrm{y}\nonumber
\end{equation} 
As it is a standard approach in \BBz literature, we also present the 95\% C.L. limit on $\tau_{1/2}^{0\nu }$ including systematic uncertainties:
\begin{equation}
\tau_{1/2}^{0\nu }\geq 2.3\times10^{24} \mathrm{y}\nonumber
\end{equation}

\section{Conclusion}
In this paper we have presented the CUORICINO final result on \BBz in \tect, obtained with an exposure of 19.75 kg$\cdot$y of \tectn, including a detailed study of systematic errors for the first time. A half life limit of 2.8 10$^{24}$~y at 90\%~C.L. is obtained (2.9 10$^{24}$~y if systematic errors are not included), to be compared (as discussed in section~\ref{sec:toy}) with an experimental sensitivity\footnote{This is the sensitivity evaluated for the total (RUN I + RUN II) statistics.} of \ca 2.6~10$^{24}$~y. This limit can be used to extract an upper limit on $m_{ee}$ using the theoretical NME evaluation for \tect nucleus. We report here results obtained using the most recent nuclear calculations found in literature:
\begin{itemize}
\item 300--570 meV using the Quasiparticle Random Phase Approximation (QRPA) evaluations of reference~\cite{NME1}
\item 360--580 meV using the QRPA evaluations of reference~\cite{NME2}
\item 570--710 meV using the Shell Model (SM) evaluations of reference~\cite{NME3}
\item 350--370 meV using the Interacting Boson Model (IBM) evaluations of reference~\cite{NME4}
\end{itemize}
\noindent Note that, for each reference, a range (and not a single value) for $m_{ee}$ is presented, reflecting the different results for the NME obtained by the authors when varying model parameters, such as the treatment of the short range correlations or the value of g$_A$ (the axial-vector coupling). Then, the interval 300--710 meV can be taken as the final range for the 90\% C.L. upper bound on  $m_{ee}$ (at 95\% C.L. this becomes 340--780 meV).

In table~\ref{table:NME} we compare this result with the most stringent 90\%~C.L. half-life lower limits present in literature. For each experimental result we report the $m_{ee}$ range obtained with the NME evaluations here considered. Despite the differences between the NME evaluations, it is evident that CUORICINO is one of the most sensitive experiments performed to date.

Finally the bottom two rows in table~\ref{table:NME} compare the 95\% C.L. half-life limit on \tect obtained in this work with the 2 sigma range corresponding to the positive signal quoted by ~\cite{Klapdor} and obtained with a re-analysis of the Heidelberg-Moscow data. The two results are clearly compatible.\\

\section{Acknowledgements}

The CUORICINO Collaboration owes many thanks to the Directors and Staff of the Laboratori Nazionali del Gran Sasso over the years of the development, construction and operation of CUORICINO, and to the technical staffs of our Laboratories. In particular we would like to thank R.~Gaigher, R.~Mazza, P.~Nuvolone, M.~Perego, B.~Romualdi, L.~Tatananni and A.~Rotilio for continuous and constructive help in various stages of this experiment.
We are grateful to our colleagues of the CUORE collaboration for valuable suggestions. Among them we would like especially to name Yury Kolomenski for help and fruitful discussions. 
 The experiment was supported by the Instituto Nazionale di Fisica Nucleare (INFN), the Commission of the European Community under Contract No. HPRN-CT-2002-00322, by the U.S. Department of Energy under Contract No. DE-AC03-76-SF00098, and DOE W-7405-Eng-48, and by the National Science Foundation Grant Nos. PHY-0139294 and PHY-0500337. \\

\end{document}

%% file: authors-CUORICINO.tex
\author[Como,INFNMilano]{E.~Andreotti}
\author[Milano,INFNMilano]{C.~Arnaboldi}
\author[USC]{F.~T.~Avignone~III}
\author[LNGS]{M.~Balata}
\author[USC]{I.~Bandac}
\author[Firenze]{M.~Barucci}
\author[LBNL]{J.~W.~Beeman}
\author[Roma]{F.~Bellini}
\author[Milano,INFNMilano]{C.~Brofferio}
\author[LBNL,BerkeleyPhys]{A.~Bryant}
\author[LNGS]{C.~Bucci}
\author[Genova,INFNGenova]{L.~Canonica}
\author[Milano,INFNMilano]{S.~Capelli}
\author[INFNMilano]{L.~Carbone}
\author[Milano,INFNMilano]{M.~Carrettoni}
\author[Milano,INFNMilano]{M.~Clemenza}
\author[INFNMilano]{O.~Cremonesi}
\author[USC]{R.~J.~Creswick}
\author[Genova,INFNGenova]{S.~Di~Domizio}
\author[LLNL,BerkeleyPhys]{M.~J.~Dolinski}
\author[Wisc]{L.~Ejzak}
\author[Roma]{R.~Faccini}
\author[USC]{H.~A.~Farach}
\author[Milano,INFNMilano]{E.~Ferri}
\author[Milano,INFNMilano]{E.~Fiorini}
\author[Como,INFNMilano]{L.~Foggetta}
\author[INFNMilano]{A.~Giachero}
\author[Milano,INFNMilano]{L.~Gironi}
\author[Como,INFNMilano]{A.~Giuliani}
\author[LNGS]{P.~Gorla}
\author[LNGS,INFNGenova]{E.~Guardincerri}
\author[CalPoly]{T.~D.~Gutierrez}
\author[LBNL,BerkeleyMat]{E.~E.~Haller}
\author[LLNL]{K.~Kazkaz}
\author[Milano,INFNMilano]{S.~Kraft}
\author[LBNL,BerkeleyPhys]{L.~Kogler}
\author[Milano,INFNMilano]{C.~Maiano}
\author[Wisc]{R.~H.~Maruyama}
\author[USC]{C.~Martinez}
\author[INFNMilano]{M.~Martinez}
\author[USC]{S.~Newman}
\author[LNGS]{S.~Nisi}
\author[Como,INFNMilano]{C.~Nones}
\author[LLNL,BerkeleyNuc]{E.~B.~Norman}
\author[Milano,INFNMilano]{A.~Nucciotti}
\author[Roma]{F.~Orio}
\author[Genova,INFNGenova]{M.~Pallavicini}
\author[Legnaro]{V.~Palmieri}
\author[Milano,INFNMilano]{L.~Pattavina}
\author[Milano,INFNMilano]{M.~Pavan}
\author[LLNL]{M.~Pedretti}
\author[INFNMilano]{G.~Pessina}
\author[INFNMilano]{S.~Pirro}
\author[INFNMilano]{E.~Previtali}
\author[Firenze]{L.~Risegari}
\author[USC]{C.~Rosenfeld}
\author[Como,INFNMilano]{C.~Rusconi}
\author[Como,INFNMilano]{C.~Salvioni}
\author[LLNL]{S.~Sangiorgio}
\author[Milano,INFNMilano]{D.~Schaeffer}
\author[LLNL]{N.~D.~Scielzo}
\author[Milano,INFNMilano]{M.~Sisti}
\author[LBNL]{A.~R.~Smith}
\author[Roma]{C.~Tomei}
\author[Firenze]{G.~Ventura}
\author[Roma]{M.~Vignati}

  \address[Como]{Dip.\ di Fisica e Matematica dell'Univ.\ dell'Insubria and Sez.\ INFN di Milano, Como I-22100 - Italy}
  \address[Milano]{Dip.\ di Fisica dell'Universit\`{a} di Milano-Bicocca I-20126 - Italy}
  \address[INFNMilano]{Sez.\ INFN di Mi-Bicocca, Milano I-20126 -Italy}
  \address[USC]{Dept.\ of Phys.\ and Astron., Univ.\ of South Carolina, Columbia, SC 29208 - USA}
  \address[LNGS]{Laboratori Nazionali del Gran Sasso, I-67010, Assergi (L'Aquila) - Italy}
  \address[Firenze]{Dip.\ di Fisica dell'Universit\`{a} di Firenze and Sez.\ INFN di Firenze, Firenze I-50125 - Italy}
  \address[LBNL]{Lawrence Berkeley National Laboratory, Berkeley, CA 94720 - USA}
  \address[Roma]{Dip.\ di Fisica dell'Universit\`{a} di Roma La Sapienza and Sez.\ INFN di Roma, Roma  I-00185 - Italy}
  \address[BerkeleyPhys]{Dept.\ of Physics, Univ.\ of California, Berkeley, CA 94720 - USA}
  \address[Genova]{Dip.\ di Fisica dell'Universit\`{a} di Genova - Italy}
  \address[INFNGenova]{Sez.\ INFN di Genova, Genova I-16146 - Italy}
  \address[LLNL]{Lawrence Livermore National Laboratory, Livermore, CA 94550 - USA}
  \address[Wisc]{Univ.\ of Wisconsin, Madison, WI - USA}
  \address[CalPoly]{California Polytechnic State Univ., San Luis Obispo, CA 93407 - USA}
  \address[BerkeleyMat]{Dept.\ of Materials Sc.\ and Engin., Univ.\ of California, Berkeley, CA 94720 - USA}
  \address[BerkeleyNuc]{Dept.\ of Nuclear Engineering, Univ.\ of California, Berkeley, CA 94720 - USA}
  \address[Legnaro]{Laboratori Nazionali di Legnaro, I-35020 Legnaro (Padova) - Italy}